\documentclass{article}
\usepackage{spconf,amsmath,graphicx,amssymb}
\usepackage{float}

\title{Analysis vs Synthesis - An Investigation of (Co)sparse Signal Models on Graphs}
\name{Madeleine S. Kotzagiannidis, Mike E. Davies\thanks{\copyright 2018 IEEE. This work was supported by the ERC project C-SENSE (ERC-ADG-2015-694888). MD is also supported by a Royal Society Wolfson Research Merit Award.}}
\address{Institute for Digital Communications, University of Edinburgh, EH9 3JL, UK\\ \{madeleine.kotzagiannidis, mike.davies\}@ed.ac.uk}
\begin{document}
\ninept

%
\maketitle
\begin{abstract}
In this work, we present a theoretical study of signals with sparse representations in the vertex domain of a graph, which is primarily motivated by the discrepancy arising from respectively adopting a synthesis and analysis view of the graph Laplacian matrix. 
Sparsity on graphs and, in particular, the characterization of the subspaces of signals which are sparse with respect to the connectivity of the graph, as induced by analysis with a suitable graph operator, remains in general an opaque concept which we aim to elucidate. By leveraging the theory of cosparsity, we present a novel (co)sparse graph Laplacian-based signal model and characterize the underlying (structured) (co)sparsity, smoothness and localization of its solution subspaces on undirected graphs, while providing more refined statements for special cases such as circulant graphs. Ultimately, we substantiate fundamental discrepancies between the cosparse analysis and sparse synthesis models in this structured setting, by demonstrating that the former constitutes a special, constrained instance of the latter.




\end{abstract}

\section{Introduction}
\label{sec:intro}
The ability to capture geometric complexity, amid an ever-growing presence of large, irregularly and complex structured data, has rendered graphs a powerful tool for arising representation and processing tasks. Simultaneously, the extension of classical signal processing concepts and tools to signals defined on graphs has created a need for a comprehensive theoretical foundation, culminating in the formation of the field of Graph Signal Processing (GSP) \cite{shu}. With its underlying theory still in its infancy, new questions and challenges are continuously emerging as a result of the complex connectivity of networks.\\
Concepts such as \textit{sparsity} and \textit{smoothness} of a signal attain complexity when evaluated with respect to the irregular domain of graph vertices and thus require refined analysis and interpretations. While previous work \cite{shu} has predominantly focused on a notion of signal smoothness on graphs, which is related to sparse representations in the graph frequency domain, i.e. bandlimitedness, its characterization in the vertex domain remains more opaque. In particular, the problem of inducing sparsity with respect to the graph connectivity via a fundamental graph difference operator ${\bf L}$, as captured via the measure $||{\bf L}{\bf x}||_0$, and thereby characterizing the classes of signals ${\bf x}$, beyond piecewise constant signals, which can be (partially) annihilated on a graph, has received less attention \cite{splinesw}. In this work, we consider graph signals, which have a sparse representation with respect to a designated graph difference operator, to be \textit{piecewise-smooth}; this notion is primarily guided by the Green's functions and/or (variational) splines of the operator \cite{splinesw} (see \cite{splines} in the classical domain). An alternative definition of piecewise-smoothness of signals on graphs  is given in  \cite{kov}.
\\
{\bf Contributions:}
The proposed theoretical analysis is primarily conducted in an effort to capture signal sparsity in the light of the connectivity of graphs and contribute to a more rigorous foundation of GSP, while simultaneously developing crucial insight for generalized signal models beyond.
Its study was initiated in a previous body of work which developed a framework for sparse graph wavelet analysis and sampling on circulant graphs and beyond \cite{splinesw}, \cite{acha2}, which, however, only offered partial characterization of the underlying signal model when the graph at hand is circulant. This work extends and upgrades the study of sparse representations on graphs, by providing an intuitive and complete characterization in the context of GSP as well as discovering more wide-ranging implications for Union of Subspaces (UoS) signal models.\\
Inspired by the study of \cite{cos}, which seeks to identify equivalencies and discrepancies between analysis- and synthesis-driven signal models, we begin by interpreting the graph Laplacian matrix ${\bf L}$ respectively in an analysis framework, and, via its Moore-Penrose Pseudoinverse (MPP) ${\bf L}^{\dagger}$, in a synthesis framework. By distinctly characterizing and comparing the associated signal subspaces, we discover how the singularity of ${\bf L}$ creates a discrepancy in the type and localization of the underlying piecewise-smooth signal classes of each model as well as dictates their associated (structured) sparsity pattern. Subsequently, we concretely demonstrate for circulant graphs, that while the sparse synthesis model induces up to piecewise quadratic polynomial signals, the cosparse analysis model is degree-reduced and only encompasses up to piecewise linear polynomials, both of which are subject to graph dependent perturbations. 
\\
{\bf Related Work:}
While GSP has featured both synthesis and analysis driven approaches, ranging from dictionary design and learning \cite{dict} to (multiresolution) graph wavelet analysis and filterbank construction (e.g. \cite{Coifman}, \cite{spectral}, \cite{ortega3}), in certain cases one inducing the other \cite{kov}, \cite{tools}, a comparative study of the two has, to the best of our knowledge, not yet been realized.
In \cite{Pesenson}, variational splines on graphs were defined as the Green's functions of a regularized graph Laplacian operator, however, due to its invertibility, the operator loses the distinctive features of ${\bf L}$, while the underlying signal model constitutes an approximation on the graph.
\\
We begin by stating preliminaries in Sect. $2$ and the analysis vs synthesis problem in Sect. $3$, before presenting the (co)sparse graph signal models in Sect. $4$. Further, in Sect. $5$, we conduct a case study on circulant graphs and present illustrative examples, with concluding remarks in Sect. $6$. We omit proofs and extensions for brevity, which will be included in an upcoming comprehensive publication.
\section{Preliminaries}
A graph $G=(V,E)$ is defined by a set $V=\{0,...,N-1\}$ of vertices of cardinality $|V|=N$ and a set $E$ of edges, which connect pairs of vertices. The adjacency matrix ${\bf A}\in\mathbb{R}^{N\times N}$ of $G$ captures its connectivity by assigning non-zero weights to existing edges between pairs of vertices $\{i,j\}$ at entries $A_{i,j}>0$ and  $A_{i,j}=0$ otherwise, while the diagonal degree matrix ${\bf D}$ with $D_{i,i}=\sum_j A_{i,j}$ contains the degree at each vertex. The (non-normalized) graph Laplacian, given by ${\bf L}={\bf D}-{\bf A}$, constitutes a fundamental graph difference matrix. In this work, we a priori assume that the graph at hand is both undirected and connected. In particular, ${\bf L}$ is symmetric positive semi-definite with a complete set of orthonormal eigenvectors $\{{\bf u}_l\}_{l=0}^{N-1}$ and nonnegative eigenvalues $0=\lambda_0< \lambda_1\leq ..\leq \lambda_{N-1}$. The operator ${\bf L}^k, \enskip k\in\mathbb{N}$, is strictly $k$-hop localized in the vertex domain, with $({\bf L}^k)_{i,j}=0$ when the shortest-path distance between $i,j$ is greater than $k$.\\
Circulant graphs, denoted with $G_S$, are defined with respect to a generating set $S=\{s_k\}_{k=1}^{M}$, $0< s_k\leq N/2$, with elements indicating edges between vertex pairs $(i,(i\pm s_k)_{modN})$ such that there exists a labelling for which ${\bf L}$ is circulant. 
The symmetric circulant matrix ${\bf L}$, with first row $\lbrack l_0\enskip l_1\enskip l_2 ... l_2\enskip l_1\rbrack$ and bandwidth $M$, can be defined via its representer polynomial $l(z) = l_0 +  \sum^M_{i=1} l_i(z^i + z^{-i})$.\\
A graph signal on $G$ of dimension $N$ is a complex-valued scalar function, which assigns the sample value $x(i)$ to node $i$ and is represented as the vector ${\bf x}\in\mathbb{C}^N$. We define ${\bf x}$ to be piecewise-smooth on $G$ with respect to ${\bf L}$ if its representation ${\bf L}{\bf x}$ is sparse, i.e. $||{\bf L}{\bf x}||_0\ll N$. Further, we define (piecewise) polynomials on $G$ such that, for a given labelling, the corresponding ${\bf x}$ is the discrete, vectorized version of a classical (piecewise) polynomial (see Def. $3.1$, \cite{splinesw}). 
The oriented edge-vertex incidence matrix ${\bf S}\in\mathbb{R}^{|E|\times N}$ of $G$, assigns an arbitrary but fixed direction to each edge, 
with $S_{k,i}=\sqrt{A_{i,j}}$ and $S_{k,j}=-\sqrt{A_{i,j}}$ if the $k$-th edge $\{i,j\}$ is directed from $i$ to $j$,   
resulting in the operation $({\bf S}{\bf x})_{\{i,j\}}=\sqrt{A_{i,j}}(x(i)-x(j))$ at edge $\{i,j\}$. Analogously to discrete differential geometry operators and as outlined in prior work \cite{splinesw}, ${\bf L}$ constitutes a graph-realization of a second-order differential operator, while ${\bf S}$ is linked to a first-order differential operator. Moreover, the matrices are related via ${\bf L}={\bf S}^T{\bf S}$ and have rank $N-k$ for $k$ connected components in $G$. When the graph is connected with $k=1$, we have $N({\bf L})=N({\bf S})=z{\bf 1}_N, \enskip z\in\mathbb{R}$, where $N(\cdot)$ denotes the nullspace.




\subsection{Notation}
The sparsity of ${\bf x}$ is given by $||{\bf x}||_0=\#\{i:x_i\neq0\}$. The canonical basis vectors ${\bf e}_i$ satisfy $e_i(i)=1$ and $e_i(j)=0,\enskip j\neq i$, and ${\bf J}_N$ is the all-ones matrix of size $N$. Let ${\bf \Psi}_{\Lambda}$ be a sampling matrix with $\Psi_{\Lambda}( i,j)=1$ for $j=\Lambda_i\in \Lambda$
and $0$ otherwise, for ordered index set $\Lambda\subset \lbrack 0\enskip... N-1\rbrack$, where $\Lambda_i$ is the $i$-th element in $\Lambda$, and let ${\bf \Psi}_{\Lambda}{\bf x}={\bf x}_{ \Lambda}$. In the synthesis model, ${\bf D}_{\Lambda}$ denotes the ${\Lambda}$-indexed columns of ${\bf D}$, while for analysis, ${\bf \Omega}_{\Lambda}$ represents the rows ${\Lambda}$ of ${\bf \Omega}$.

\section{Analysis vs Synthesis Signal Models}
The synthesis (or generative) model encompasses signals which can be described through few linear combinations of the atoms of a dictionary ${\bf D}\in\mathbb{R}^{N\times M}$, with $M\geq N$, i.e. there exists a ${\bf c}\in\mathbb{R}^{M}$ with small $k=||{\bf c}||_0$ so that ${\bf x}={\bf D}{\bf c}$. In contrast, the analysis model implicitly describes the sparsity of representation ${\bf x}$ through an analysis operator ${\bf \Omega}\in\mathbb{R}^{M\times N}$ with $||{\bf \Omega}{\bf x}||_0=k$. In particular, this model has been shown to promote the concept of \textit{cosparsity}, where the emphasis is placed on the number of zeros $l=M-k$ (cosparsity), as opposed to the number of non-zero coefficients \cite{cos}. By letting $\Lambda$ and $\Lambda^{\complement}$ denote the locations of the cosparse and sparse entries respectively such that ${\bf \Omega}_{\Lambda}{\bf x}={\bf 0}_{\Lambda}$ and ${\bf x}={\bf D}_{\Lambda^{\complement}}{\bf c}_{\Lambda^{\complement}}$, we may describe their respective solution subspaces as ${W}_{\Lambda}:=N({\bf \Omega}_{\Lambda})$ and  ${V}_{\Lambda^{\complement}}:=span({\bf D}_j,j\in\Lambda^{\complement})$. As such, both models represent instances of the more general UoS model \cite{cos}, yet they are inherently different.  
When the operators are trivially related via an inverse operation ${\bf D}={\bf \Omega}^{-1}$ the two models become equivalent, however, when the operators are rank-deficient and/or rectangular, discrepancies arise, whose study has only recently been initiated (\cite{cos}, \cite{elad}).

\section{(Co)sparse graph signal models}
In this study, we primarily focus on the graph Laplacian as an analysis operator ${\bf \Omega}={\bf L}$ in the vertex domain of a graph $G$ and define a synthesis counterpart through its MPP ${\bf D}={\bf L}^{\dagger}$, representing a dictionary of graph signals.
\subsection{Analysis}
In order to completely characterize the types of signals ${\bf x}$, defined on the vertices of a connected graph, which can be (partially) annihilated by the corresponding graph Laplacian, we need to consider the subspaces induced by $N({\bf \Psi}_{\Lambda}{\bf L})$, where ${\Lambda}\subset V$ is an arbitrary \textit{cosupport} associated with the zeros of ${\bf L}{\bf x}$. We thus state the following:\\
\\
\textit{{\bf Proposition 1.} The nullspace of ${\bf \Psi}_{\Lambda}{\bf L}$, where ${\bf L}$ is the graph Laplacian of an undirected connected graph $G=(V,E)$, has rank $|\Lambda^{\complement}|$ and is described by $N({\bf \Psi}_{\Lambda}{\bf L})=z{\bf 1}_N +{\bf L}^{\dagger}{\bf \Psi}_{\Lambda^{\complement}}^T{\bf W}{\bf c}$, for arbitrary $z\in\mathbb{R}$, ${\bf c}\in\mathbb{R}^{|\Lambda^{\complement}|-1}$, and ${\bf W}\in\mathbb{R}^{|\Lambda^{\complement}|\times(|\Lambda^{\complement}|-1)}$ given by
 \begin{equation}{\bf W}:=\scalebox{0.8}{$\begin{pmatrix} |\Lambda^{\complement}|-1&0&&\dots& &0\\
-1&|\Lambda^{\complement}|-2& 0&\dots& &0\\
&-1&|\Lambda^{\complement}|-3&&&\vdots\\
\vdots&&&&&\\
&&&&&0\\&&&&&1\\-1&-1&\dots&&&-1\end{pmatrix}$}.\end{equation}
}
\newline \textit{Remark 1}: The zero-sum column structure of ${\bf W}$ mirrors that of transposed incidence matrix ${\bf S}^T$, up to a weight $\sqrt{A_{i,j}}$ per column, and we can alternatively express ${\bf \Psi}_{\Lambda^{\complement}}^T{\bf W}$ as a basis in $({\bf e}_i-{\bf e}_j)$, for any $i,j\in {\Lambda^{\complement}}\subset V$ which are connected via a path. 
\\
\\
It becomes evident that Prop.\ 1 offers a synthesis description of the analysis model. We further note that, provided $|\Lambda|<N$, the matrix ${\bf \Psi}_{\Lambda}{\bf L}$ always has full row-rank, while $N({\bf \Psi}_{\Lambda}{\bf L})$ consists of the space spanned by $N({\bf L})$ of rank $1$ and ${\bf L}^{\dagger}{\bf \Psi}_{\Lambda^{\complement}}^T{\bf W}$ of rank $|{\Lambda}^{\complement}|-1$, such that their composition $N({\bf \Psi}_{\Lambda}{\bf L})$ has rank $N-|{\Lambda}|=|{\Lambda}^{\complement}|$. \\
\\
\textit{Remark 2}: When the graph is disconnected, this analysis can be adapted to each connected component and gives rise to a block-wise signal model, which here is omitted for brevity.


\subsection{Synthesis}
Consider arbitrary sparse graph signal ${\bf c}\in\mathbb{R}^N$ with support ${\Lambda^{\complement}}\subset V$ and representation ${\bf x}={\bf L}^{\dagger}{\bf c}={\bf L}^{\dagger}_{\Lambda^{\complement}}{\bf c}_{\Lambda^{\complement}}$ on a connected graph $G$, via graph synthesis operator ${\bf L}^{\dagger}$. In the following, we further characterize this synthesis representation via its relation to the analysis operator ${\bf L}$.
We begin by noting the MPP relations ${\bf L}{\bf L}^{\dagger}={\bf I}_N-\frac{1}{N}{\bf J}_N$, which constitutes the projection onto the subspace $N({\bf L})^{\perp}$, and ${\bf L}{\bf L}^{\dagger}{\bf L}={\bf L}$. Due to the Gramian structure of ${\bf L}$, this further implies:  
\newline {\bf Discontinuity Property}.\ We have ${\bf L}({\bf L}^{\dagger}{\bf S}^T)={\bf S}^T$, which reveals that the possible locations of the discontinuities (knots) of the piecewise-smooth functions $({\bf L}^{\dagger}{\bf S}^T)_j={\bf S}^{\dagger}_j$ with respect to ${\bf L}$ are in the space spanned by the $2$-sparse columns of ${\bf S}^T$, extending Rem.\ $1$.\\
\\
More generally, we deduce that any graph signal that is sparse with respect to ${\bf L}$ must have a distinctly \textit{structured sparsity} pattern in the range of the columns ${\bf S}_j^T$: if the graph is connected, the locations of non-zeros are arbitrary with the constraint that the sum of their values is $0$.
Further, we discover from the relation ${\bf L}{\bf L}^{\dagger}{\bf L}={\bf L}^2{\bf L}^{\dagger}={\bf L}$ that when the graph is sufficiently sparse, the non-zero entries of the columns $({\bf L})_j$ can be interpreted as the locations of the knots of $({\bf L}^{\dagger})_j=({\bf L}^{\dagger 2}{\bf L})_j$ with respect to the higher-order and $2$-hop localized operator ${\bf L}^2$. Overall, this reveals that a signal which is analysis-sparse defines a restricted synthesis representation. 
\\
The synthesis representation ${\bf x}={\bf L}^{\dagger}{\bf c}$ is generally not sparse with respect to ${\bf L}$ unless ${\bf c}$ satisfies the above constraints. Further, we have ${\bf L}^2{\bf x}={\bf L}{\bf c}$ which is sparse only if ${\bf L}$ is sufficiently sparse, whereby the non-zeros of ${\bf L}{\bf c}$ correspond to the knot locations of ${\bf L}^{\dagger}{\bf c}={\bf L}^{\dagger ^2}{\bf L}{\bf c}$ with respect to ${\bf L}^2$.\\
\\
We observe that the underlying graph signal spaces induced by respectively the (co)sparse analysis and synthesis models vary in degree of smoothness and graph localization:\\
\\
{\bf Localization and Smoothness}
{\bf Property}: ${\bf L}^{\dagger}$ and ${\bf S}^{\dagger}$ respectively represent the Green's functions (matrices) of ${\bf L}$ and ${\bf S}$, and as such capture different levels of smoothness. The functions ${\bf S}_j^{\dagger}$ are piecewise smooth with respect to ${\bf L}$ which induces sparsity localized with respect to a 1-hop neighborhood of the graph. For sufficiently sparse graphs, the functions ${\bf L}_j^{\dagger}$ are piecewise smooth with respect to ${\bf L}^2$ which induces 2-hop localized sparsity.\\
\\
Overall, we conclude:\\
\textit{{\bf Theorem 1.} The cosparse analysis model, described by $N({\bf \Psi}_{\Lambda}{\bf L})$, constitutes a constrained case of the sparse synthesis model, generated by ${\bf L}^{\dagger}{\bf \Psi}_{\Lambda^{\complement}}^T$, with respect to matrix ${\bf W}$ (Eq.\ $(1)$) which satisfies $\langle{\bf \Psi}_{\Lambda^{\complement}}^T{\bf W}{\bf e}_j{,}N({\bf L})\rangle=0\ \forall j$, up to a shift by $N({\bf L})=z{\bf 1}_N,\ z\in\mathbb{R}$.} 









\subsection{Uniqueness}
Let ${\bf M}\in\mathbb{R}^{m\times N}$, $m<N$, denote a suitable (graph-)measurement matrix with linearly independent rows and consider the problem of identifying the unique (co)sparse solution of ${\bf y}={\bf M}{\bf x}$. In the synthesis model, we require that $2k<spark({\bf M D })\leq m+1$, in order to uniquely identify the unknown $k$-sparse signal ${\bf c}$ \cite{cos}. In the analysis model, an equivalent measure to the spark, $\kappa_{{\bf \Omega}}(l):=\max_{|\Lambda|\geq l}dim(W_{\Lambda})$ is established and for mutually independent ${\bf M}$ and ${\bf \Omega}$ (i.e. their rows do not have non-trivial linear dependencies) we require $2\kappa_{{\bf \Omega}}(l)\leq m$ to determine the $l$-cosparse signal ${\bf x}$ (see Prop. $3$, \cite{cos}). Due to its uniquely characterizable linear dependencies and following Prop. $1$, these interdependency measures for ${\bf L}$ are identified as $spark({\bf L}^{\dagger})=N$ and $\kappa_{{\bf L}}(l)=N-l=k$ for $l<N$, facilitating the following result for the analysis model:\\
\\
\textit{{\bf Corollary 1.} For mutually independent ${\bf M}$ and ${\bf L}$, the graph-Laplacian-based problem
\[{\bf M}{\bf x}={\bf y}\enskip\text{with}\enskip ||{\bf L}{\bf x}||_0\leq N-l  \] 
 has at most one solution if $m\geq 2(N-l)$.}
	


\section{(Co)sparsity on Circulant Graphs}
\subsection{Signal Models}
In previous work, it has been shown that the circulant graph Laplacian ${\bf L}$ annihilates up to linear polynomial signals, subject to a border effect dependent on the bandwidth of the graph, i.e. its representer polynomial $l(z)$ has $2$ vanishing moments (Lemma $3.1$, \cite{splinesw}). \\
Further, the simple cycle $G_S$, with $S=\{1\}$ and representer polynomial $l_C(z)=2-z^{-1}-z$, as the base case of circulant graphs with a direct analogy to discrete time-periodic signal processing, possesses a uniquely characterizable graph Laplacian MPP:\\
\\
\textit{{\bf Property}} (see Thm. $1$, \cite{green2}). The pseudoinverse ${\bf L}_C^{\dagger}$ of the (unnormalized) graph Laplacian of the simple cycle has entries
\newline $L_C^{\dagger}(i,j)=\frac{(N-1)(N+1)}{12N}-\frac{1}{2}|j-i|+\frac{(j-i)^2}{2N},\enskip\text{for}\enskip 0\leq i,j\leq N-1.$\\
\\
In particular, we observe that the rows and columns of  ${\bf L}_C^{\dagger}$ describe the sum of piecewise quadratic and linear polynomial terms. It becomes evident that taking unweighted differences between any pair of columns/rows produces piecewise linear signals, eliminating the quadratic term, which can be used to re-derive the annihilation property of ${\bf L}_C$, and, more generally, to define the space $N({\bf \Psi}_{\Lambda}{\bf L}_C)$. Extending (Lemma $3.1$, \cite{splinesw}), we derive the following property:\\
\\
\textit{{\bf Lemma 1}. The graph Laplacian ${\bf L}$ of a circulant graph $G_S=(V,E)$, with generating set $S$ such that $s=1\in S$ and bandwidth $M<N/2$, can be decomposed as ${\bf L}={\bf P}_{G_S}{\bf L}_C$, where ${\bf P}_{G_S}$ is a circulant positive definite matrix of bandwidth $M-1$, which depends on $G_S$ with representer polynomial 
\[P_{G_S}(z)= \left(\sum_{i=1}^{M} i d_i \right)+\sum_{i=1}^{M-1} \left(\sum_{k=i+1}^{M}(k-i) d_k \right) (z^i+z^{-i})\]
and weights $d_i=A_{j,(i+j)_{modN}}$, and ${\bf L}_C$ denotes the graph Laplacian of the simple cycle.}\\
Here, $s=1\in S$ is a sufficient condition and ensures that $G_S$ is connected. Leveraging this decomposition of ${\bf L}$ on circulant graphs, we further establish:\\
\\
\textit{{\bf Lemma 2}. The MPP of ${\bf L}$ in Lemma $1$ can be decomposed as ${\bf L}^{\dagger}={\bf P}_{G_S}^{-1}{\bf L}^{\dagger}_C$ and its rows/columns constitute piecewise quadratic polynomials which are `perturbed' by ${\bf P}_{G_S}^{-1}$.}\\
\\
\textit{Remark 3}. According to \cite{volkov}, the inverse of a cyclically banded positive definite matrix, such as ${\bf P}_{G_S}$ for banded $G_S$, contains entries that decay exponentially (in absolute value) away from the diagonal and corners of the matrix, invoking a `perturbation' on ${\bf L}^{\dagger}_C$.\\ 
\\
In addition, we can infer the following property for the elementary piecewise-smooth functions $({\bf S}^{\dagger})_j$ on $G_S$:\\
\\
{\bf Lemma 3}. \textit{The columns of ${\bf S}^{\dagger}$ for a circulant graph (as above) are piecewise linear polynomials, subject to a perturbation by ${\bf P}_{G_S}^{-1}$.}\\
\\
In light of the above, we proceed to state a concrete discrepancy between the analysis and synthesis models:\\
\\
{\bf Theorem 2}. \textit{On circulant graphs (as per Lemma $1$), the signal subspaces associated with the cosparse analysis model of ${\bf L}$ consist of up to piecewise linear polynomials, subject to a shift by $N({\bf L})$, while those of the sparse synthesis model with ${\bf L}^{\dagger}$ describe up to piecewise quadratic polynomials, respectively subject to graph-dependent perturbations in form of factor ${\bf P}^{-1}_{G_S}$.}\\
\\
The synthesis model defines graph signals as linear combinations of (perturbed) piecewise quadratic polynomials, encompassing piecewise linear and piecewise constant signals for suitable coefficient choices. This signifies a more comprehensive model containing different orders of smoothness and localization of the solution spaces, while the analysis model constitutes a structured instance thereof which defines signals that are sparse with respect to ${\bf L}$, up to a shift by $N({\bf L})$. For sufficiently banded $G_S$, the `perturbed' quadratic $({\bf L}^{\dagger})_j$ can further be annihilated by ${\bf L}^2$, whose rows, by extension, factor $4$ vanishing moments, as per (Lemma $3.1$, \cite{splinesw}).\\
\begin{figure}
\begin{minipage}[b]{0.99\linewidth}
  \centering
  \centerline{\includegraphics[width=9.5cm]{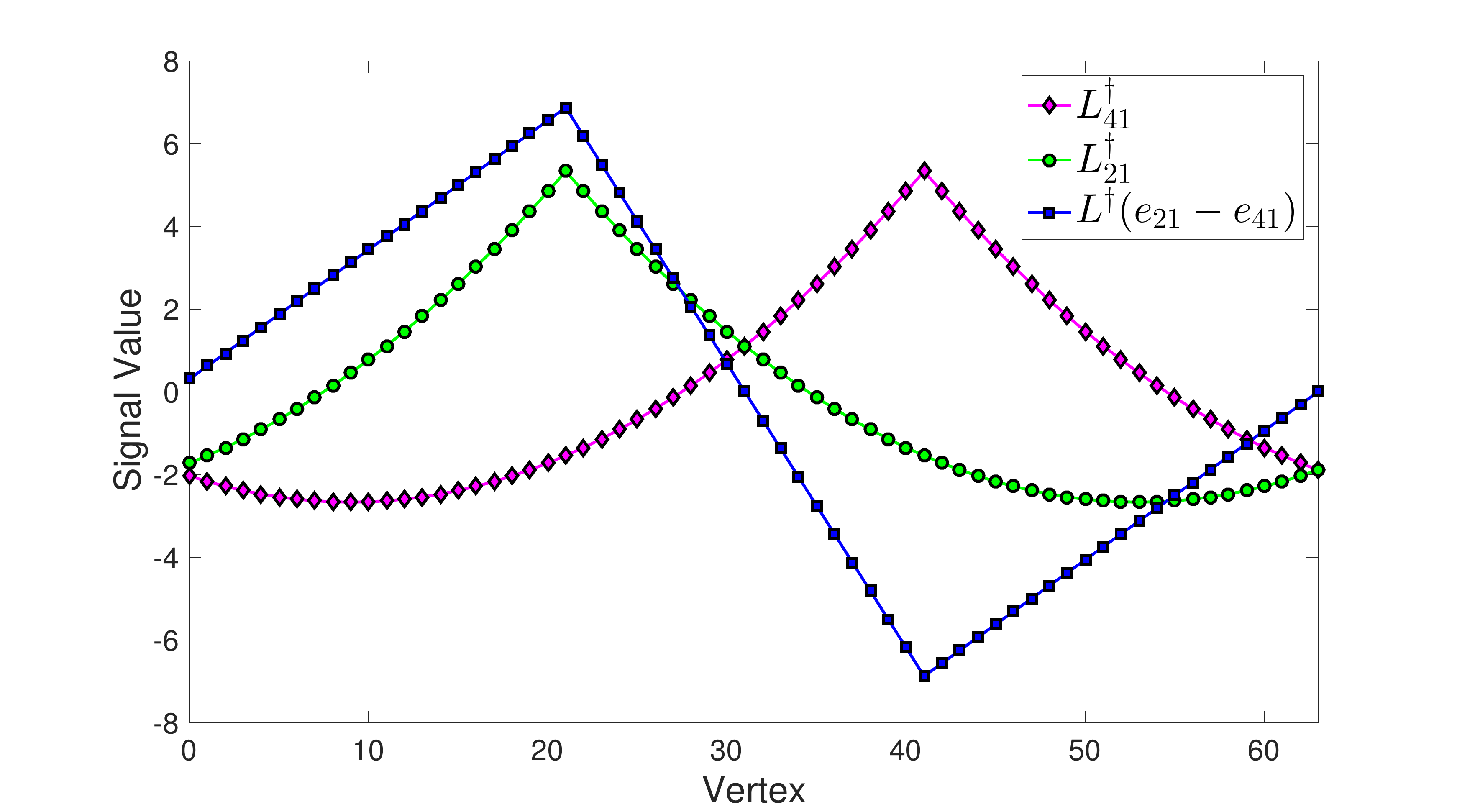}}
  \centerline{(a) PCW-Smooth Functions on $G_S$ with $S=\{1\}$}\medskip
\end{minipage}
\vfill
\begin{minipage}[b]{0.99\linewidth}
  \centering
  \centerline{\includegraphics[width=9.5cm]{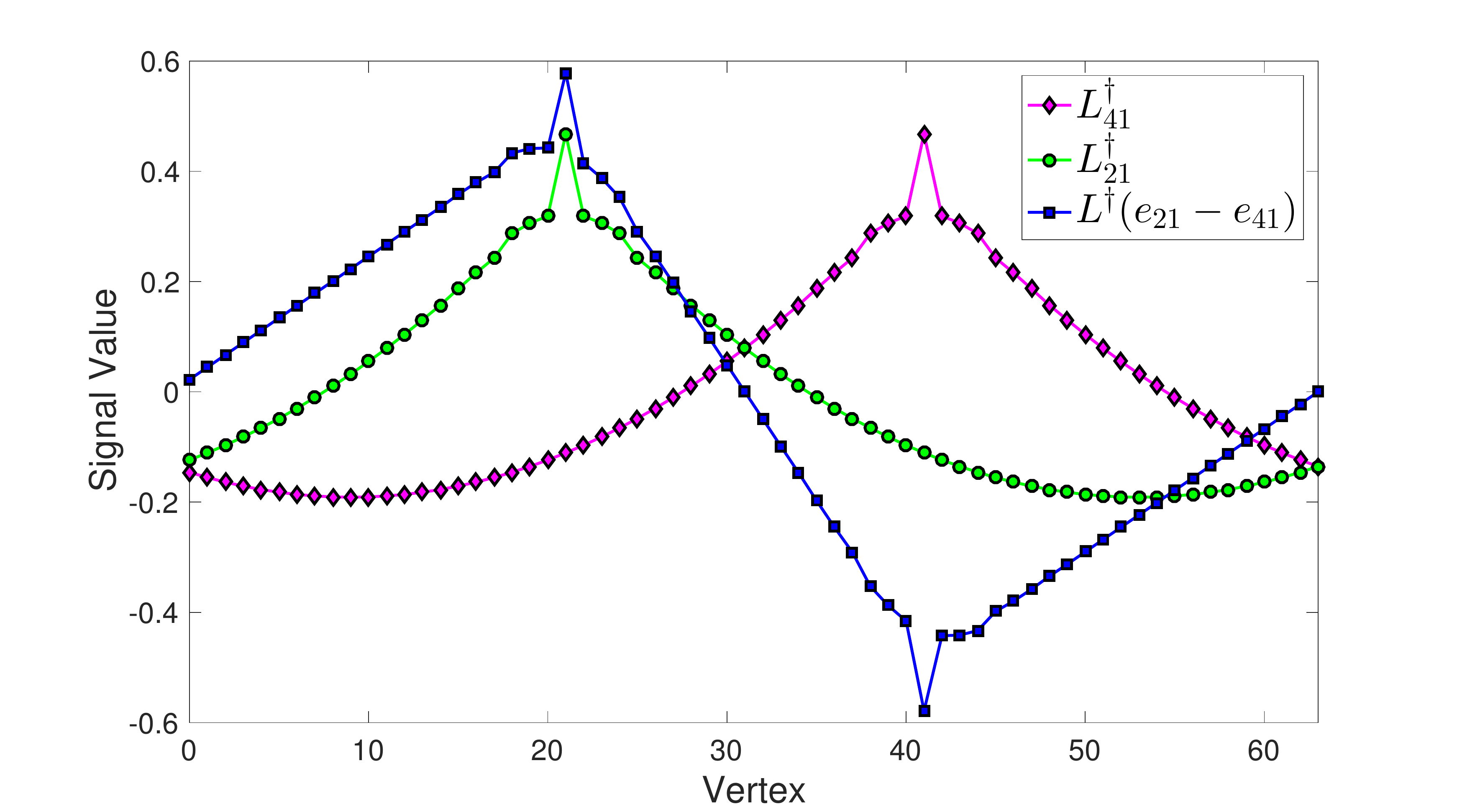}}
  \centerline{(b) PCW-Smooth Functions on $G_S$ with $S=\{1,2,3\}$}\medskip
\end{minipage}
\begin{minipage}[b]{0.99\linewidth}
  \centering
  \centerline{\includegraphics[width=9cm]{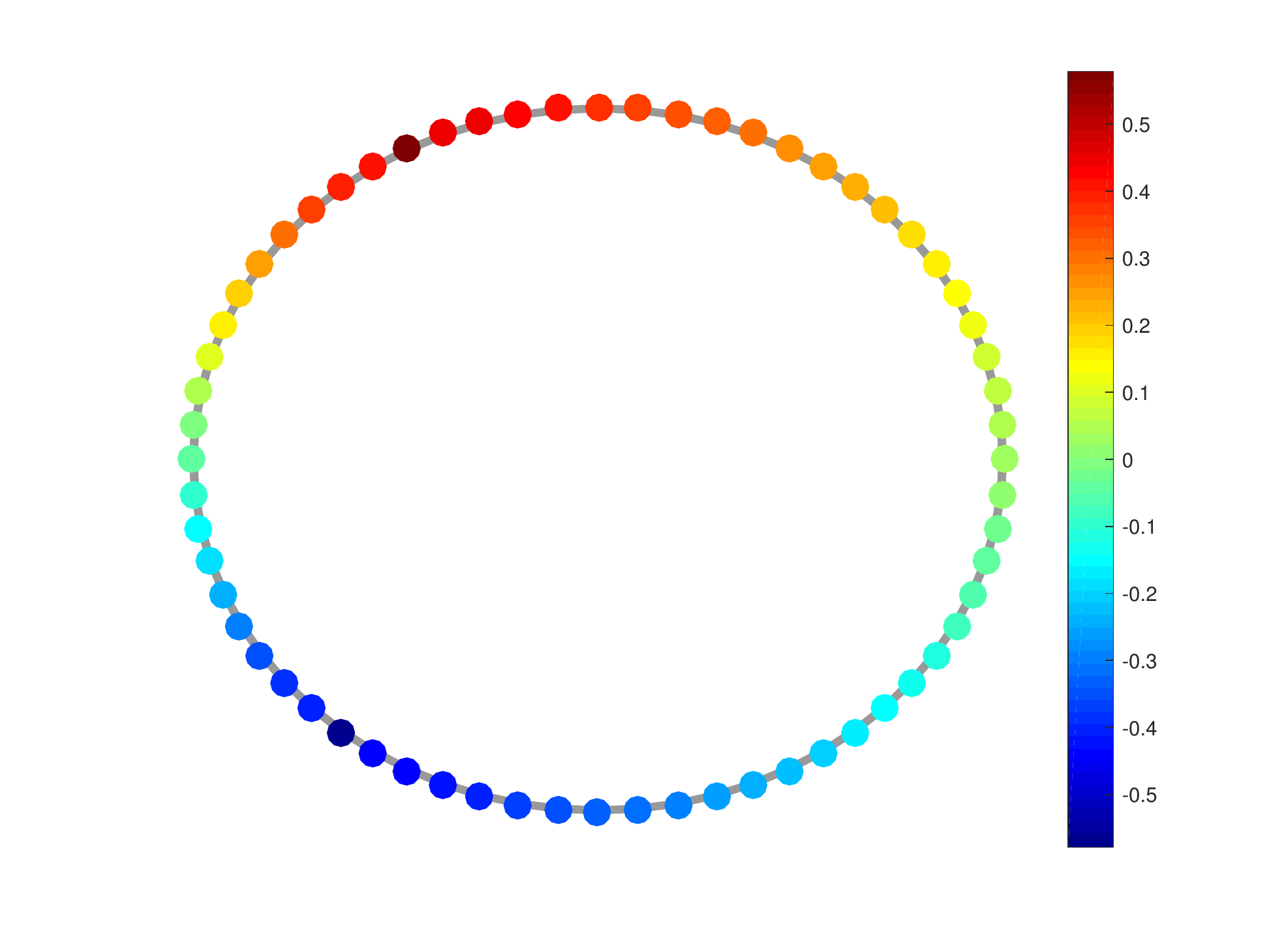}}
  \centerline{(c) ${\bf L}^{\dagger}({\bf e}_{21}-{\bf e}_{41})$ on $G_S$ with $S=\{1,2,3\}$}\medskip
\end{minipage}
\caption{Comparison of signal models on circulant graphs}
\label{fig:res}
\end{figure}
\\
{\bf Example (Complete Graph)}. In the special case of an unweighted complete (circulant) graph, we discover ${\bf S}^{\dagger}=\frac{1}{{N}}{\bf S}^{T}$ and ${\bf L}^{\dagger}=\frac{1}{{N^2}}{\bf L}$. Here, the representation $({\bf S}^{\dagger})_j$ is trivially smooth with respect to ${\bf L}$, as it simultaneously represents its sparsity pattern. Nevertheless, since the graph is dense, $({\bf L}^{\dagger})_j$ is not sparse with respect to ${\bf L}^2$. Therefore, as a result of the maximum graph connectivity, we observe that the piecewise-smooth functions $({\bf S}^{\dagger})_j$ take the form of two opposite-sign impulses, as per $({\bf S}^{T})_j$, while $({\bf L}^{\dagger})_j$ is piecewise constant, just as $({\bf L})_j$.\\
\\
{\bf Discontinuities}.
We note that the piecewise-smooth (analysis) representation ${\bf x}={\bf P}_{G_S}^{-1}{\bf L}_C^{\dagger}{\bf \Psi}_{\Lambda^{\complement}}^T{\bf W}{\bf c}$, where ${\bf P}_{G_S}$ encapsulates the additional connectivity information of ${G_S}$, may assume different orders of smoothness depending on the choice of basis ${\bf W}$. Specifically, one may select a basis ${\bf \Psi}_{\Lambda^{\complement}}^T\tilde{{\bf W}}$, which absorbs column $({\bf P}_{G_S})_j$ while satisfying the constraints of ${\bf W}$ (Eq. $(1)$) via circular convolution $\ast$ modulo $N$, provided that $\Lambda^{\complement}$ covers the support of ${\bf p}:=(({\bf P}_{G_S})_j\ast({\bf e}_k-{\bf e}_l))_{mod N}$, for suitable $j,k,l\in V$. Thereby, ${\bf x}={\bf L}^{\dagger}{\bf p}$ becomes sparse with respect to ${\bf L}_C$, in addition to ${\bf L}$, with sparsity patterns respectively given by $({\bf e}_j\ast({\bf e}_k-{\bf e}_l))_{mod N}$ and ${\bf p}$, and can be considered piecewise-smooth in the `classical' sense as well as with respect to $G_S$, removing the perturbation by ${\bf P}^{-1}_{G_S}$.
%
\subsection{Illustrative Examples}
We illustrate the differences in the piecewise (PCW)-smooth solution spaces of the two models for two unweighted circulant graphs with $N=64$ in Fig.\ $1 (a)$-$(b)$. 
Here, two distinct quadratic polynomial atoms ${\bf L}_t^{\dagger}$ with one discontinuity respectively at $t=i,j$ are depicted in comparison to their difference, which gives rise to the linear signal with two discontinuities at $i,j$ and, hence, $2$-sparse analysis representation ${\bf e}_i-{\bf e}_j$ with respect to ${\bf L}$. The perturbation effect for the non-trivial circulant case in $(b)$ is clearly visible around the discontinuities. Fig. $1(c)$ further depicts the linear signal from $(b)$ in the vertex domain (prepared using \cite{toolbox}).



\section{Conclusion}
We have presented a comprehensive and descriptive model for (co)sparse signals on graphs by providing a refined characterization of the sparsity, smoothness and localization of underlying fundamental solution subspaces with respect to the graph Laplacian, with relevant implications for generalized UoS signal models. More generally, we have substantiated a fundamental discrepancy between the analysis and synthesis models on the basis of a structured matrix with uniquely characterizable linear dependencies, which is further concretized for circulant graphs. In current work, we aim to widen the study of signal models to a larger class of graph operators.







\bibliographystyle{IEEEbib}
\bibliography{bibles}
\end{document}